\documentclass[a4paper,letter]{aa}

\usepackage{graphicx,natbib}
\usepackage{txfonts}

\newcommand{\bz}{$\langle B_\mathrm{z} \rangle$}
\newcommand{\lep}{$\mu$\,Lep}
\newcommand{\kms}{km\,s$^{-1}$}

\newcommand{\figps}[1]{\resizebox{\hsize}{!}{\rotatebox{0}{\includegraphics{#1}}}}

\newcommand{\fifps}[2]{\centering\resizebox{#1}{!}{\includegraphics{#2}}}

\begin{document}

\title{
No magnetic field in the spotted HgMn star $\mu$\,Leporis%
\thanks{Based on observations collected at the European Southern Observatory, Chile (ESO programs 084.D-0338, 086.D-0240).}}

\author{O.~Kochukhov\inst{1}
   \and V.~Makaganiuk\inst{1}
   \and N.~Piskunov\inst{1}
   \and S.~V.~Jeffers\inst{2}
   \and C.~M.~Johns-Krull\inst{4}
   \and C.~U.~Keller\inst{2}
   \and M.~Rodenhuis\inst{2}
   \and F.~Snik\inst{2}
   \and H.~C.~Stempels\inst{1}
   \and J.~A.~Valenti\inst{3}}

\institute{Department Physics and Astronomy, Uppsala University, Box 516, 751 20 Uppsala, Sweden
\and
Sterrekundig Instituut, Universiteit Utrecht, Box 80000, NL-3508 TA Utrecht, The Netherlands
\and
Space Telescope Science Institute, 3700 San Martin Dr, Baltimore MD 21211, USA
\and
Department of Physics and Astronomy, Rice University, 6100 Main Street, Houston, TX 77005, USA}

\date{Received 30 August 2011 / Accepted 30 September 2011}

\titlerunning{No magnetic field in the spotted HgMn star $\mu$\,Leporis}
\authorrunning{O. Kochukhov et al.}

  \abstract
{Chemically peculiar stars of the mercury-manganese (HgMn) type represent a new class of spotted late-B stars, in which evolving surface chemical inhomogeneities are apparently unrelated to the presence of strong magnetic fields but are produced by some hitherto unknown astrophysical mechanism.}
{The goal of this study is to perform a detailed line profile variability analysis and carry out a sensitive magnetic field search for one of the brightest HgMn stars -- \lep.}
{We acquired a set of very high-quality intensity and polarization spectra of \lep\ with the HARPSpol polarimeter. These data were analyzed with the multiline technique of least-squares deconvolution in order to extract information on the magnetic field and line profile variability.}
{Our spectra show very weak but definite variability in the lines of Sc, all Fe-peak elements represented in the spectrum of \lep, as well as Y, Sr, and Hg. Variability might also be present in the lines of Si and Mg. Anomalous profile shapes of \ion{Ti}{ii} and \ion{Y}{ii} lines suggest a dominant axisymmetric distribution of these elements. At the same time, we found no evidence of the magnetic field in \lep, with the $3\sigma$ upper limit of only 3~G for the mean longitudinal magnetic field. This is the most stringent upper limit on the possible magnetic field derived for a spotted HgMn star.}
{The very weak variability detected for many elements in the spectrum \lep\ suggests that low-contrast chemical inhomogeneities may be common in HgMn stars and that they have not been recognized until now due to the limited precision of previous spectroscopic observations and a lack of time-series data. The null result of the magnetic field search reinforces the conclusion that formation of chemical spots in HgMn stars is not magnetically driven.}


\maketitle

\section{Introduction}
\label{intro}

Mercury-manganese (HgMn) stars form a subclass of late-B chemically peculiar stars. In addition to a large overabundance of exotic chemical elements, their most extraordinary characteristic is the presence of weak line profile variation that is attributed to chemical spots \citep{adelman:2002,hubrig:2006}. Despite occasional magnetic field detections claimed for HgMn stars \citep[e.g.,][]{hubrig:2010}, all systematic attempts to find magnetic fields using the best available data yielded null results \citep{shorlin:2002,folsom:2010,auriere:2010a,makaganiuk:2011a}. The lack of strong magnetic fields distinguishes HgMn stars from the better known strongly magnetic Ap stars, which also exhibit chemical inhomogeneities. Moreover, unlike any other known type of spotted early-type stars, the topology of spots in HgMn stars evolves on a timescale from years \citep{kochukhov:2007b} to months \citep{briquet:2010}. This unusual stellar surface structure-formation phenomenon was only observed for a handful of stars and has received no satisfactory theoretical explanation so far. Additional observational studies are needed to enlarge the sample of known spotted HgMn stars, assess the extent of surface inhomogeneity for a wider range of chemical elements, and relate spots to the possible presence of weak magnetic fields and to other stellar properties.

The third brightest HgMn star, \lep\ (HR\,1702, HD\,33904, HIP\,24305), was a popular target of the model atmosphere and abundance analysis studies of mercury-manganese stars \citep{smith:1993,woolf:1999,adelman:2000b}. Its brightness and a moderate rotational velocity, $v_{\rm e}\sin i$\,=\,15.5--16.5~\kms\ \citep{dworetsky:1998,dolk:2003}, make \lep\ an ideal object for an investigation of weak line profile variability and for a high-precision magnetic field search.

This star is classified as an $\alpha^2$\,CVn-type star in the General Catalogue of Variable Stars\footnote{\tt http://www.sai.msu.su/gcvs/gcvs/} thanks to marginal photometric changes with a period close to 2~d reported by \citet{renson:1976}. Subsequent studies did not confirm this variability \citep{heck:1987,adelman:1998b}. Recently \citet{nunez:2010} mention changes in the \ion{Hg}{ii} 3984\,\AA\ line profile. But apart from this brief report, there have been no previous spectroscopic variability studies of this star. With the exception of an early low-precision polarimetric study by \citet{babcock:1958}, \lep\ has previously not been investigated with spectropolarimetry.

\begin{table*}
\caption{Journal of spectropolarimetric observations and results of the magnetic field analysis of \lep.}
\label{tab1}
\centering
\begin{tabular}{ccccccccc}
\hline\hline
Date & Stokes & HJD & $T_{\rm exp}$ (s) & $S/N$ & $S/N$ (LSD) & FAP & \bz (V) (G) & \bz (null) (G)~  \\
\hline
2010-01-07 & $IV$ & 2455204.7676 & $4\times161$ &  750 & 5700 & 0.959 & $-0.7\pm2.5$ & $5.4\pm2.5$ \\
2011-02-12 & $IV$ & 2455605.6493 & $8\times160$ & 1000 & 8600 & 0.951 & $-2.5\pm1.8$ & $0.1\pm1.9$ \\
2011-02-13 & $IV$ & 2455606.6025 & $8\times160$ & 1150 & 9200 & 0.994 & $\phantom{-}1.4\pm1.6$ & $2.7\pm1.6$ \\
2011-02-14 & $IV$ & 2455607.6491 & $8\times160$ & 1050 & 8400 & 0.948 & $-0.4\pm1.8$ & $1.6\pm1.8$ \\
2011-02-16 & $I$  & 2455609.5099 & $8\times80$  & 700 & & & & \\
2011-02-17 & $I$  & 2455610.5128 & $8\times80$  & 850 & & & & \\
\hline
\end{tabular}
\end{table*}

\citet{berghoefer:1996} recognized \lep\ as an X-ray bright object, but the spectral properties of this emission suggest an unresolved late-type, pre-main sequence companion as the origin \citep{behar:2004}. Using adaptive imaging in the infrared \citet{scholler:2010} detect a visual companion at the separation of 0\farcs3 from the primary. The K-band luminosity difference of more than 3 mag ensures that the secondary contributes negligibly to the combined radiation at optical wavelengths.

The objective of our study is to investigate the spectral variability and assess the magnetic field properties of \lep\ in the context of a recently discovered HgMn chemical spot phenomenon. Taking advantage of the brightness of this star, we attempt to reach a significantly higher precision both in the line profile analysis and in the magnetic field search than has been achieved for other spotted HgMn stars. 

The rest of this paper is organized as follows. Section~\ref{obs} describes observational data and its reduction. Magnetic field analysis is presented in Sect.~\ref{mag}. The results of our line profile variability study are given in Sect.~\ref{lpv}. The paper concludes with a summary and discussion in Sect.~\ref{disc}.

\section{Observations and data reduction}
\label{obs}

The spectra of \lep\ analyzed in this paper were obtained with the HARPSpol polarimeter \citep{snik:2011,piskunov:2011} feeding the HARPS spectrometer \citep{mayor:2003} at the ESO 3.6-m telescope in La Silla. Five observations on different nights were obtained in February 2011. Out of these spectra, three were recorded using the circular polarization analyzer and two were obtained in the non-polarimetric mode. In addition, we reanalyzed our earlier Stokes $IV$ HARPSpol observation of \lep\ obtained in January 2010 \citep{makaganiuk:2011a}. All spectra have a resolving power of $R=115\,000$ and a peak signal-to-noise ratio (S/N) of 700--1000 per 0.8~\kms\ pixel at $\lambda\approx5200$\,\AA. The information about individual observations, including the UT and Julian dates, Stokes parameters observed, exposure times, and the S/N, is given in Table~\ref{tab1}.

Our data cover the wavelength range 3780--6913~\AA\ with a small gap around 5300~\AA. Each observation of the star was split into four to eight sub-exposures, each 80--160~s long, obtained with four different orientations of the quarter-wave retarder plate relative to the beamsplitter of the circular polarimeter. The reduction was performed using the REDUCE pipeline \citep{piskunov:2002}. The Stokes $V$ parameter and the diagnostic null spectrum were deduced with the help of the ``ratio method'' described by \citet{bagnulo:2009}. Other details of the acquisition, reduction, and calibration of the HARPSpol observations of \lep\ can be found in our previous papers devoted to HgMn stars \citep{makaganiuk:2011,makaganiuk:2011a}.

\section{Magnetic field measurements}
\label{mag}

\begin{figure}[!t]
\centering
\figps{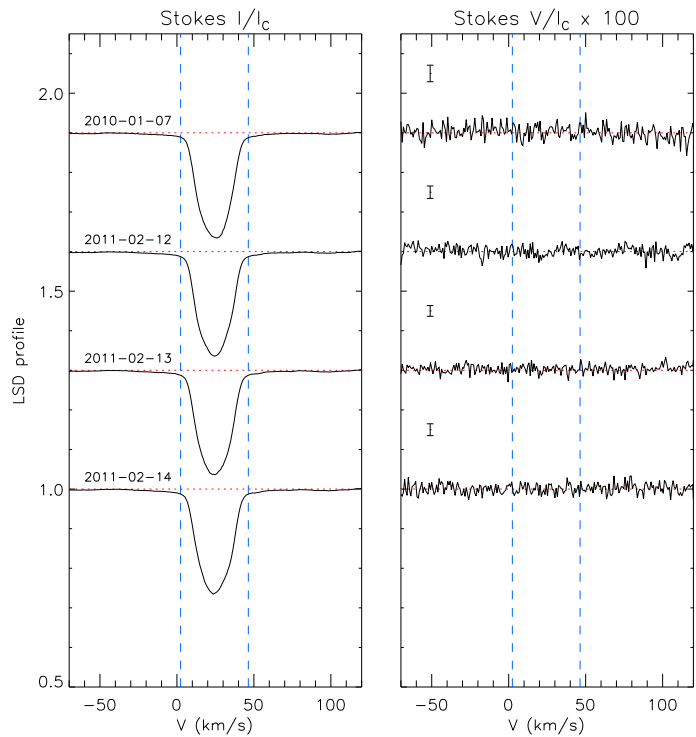}
\caption{LSD Stokes $I$ (\textit{left panel}) and Stokes $V$ (\textit{right panel}) profiles of \lep. The profiles corresponding to different observing nights are shifted vertically for clarity. The vertical scale in the Stokes $V$ panel is expanded by a factor of 100 relative to Stokes $I$. The error bars for each $V$ profile are given on the left side of the panel. The vertical dashed lines indicate the velocity range adopted for the longitudinal magnetic field measurements.}
\label{fig:lsd}
\end{figure}

\begin{figure*}[!t]
\centering
\fifps{16.5cm}{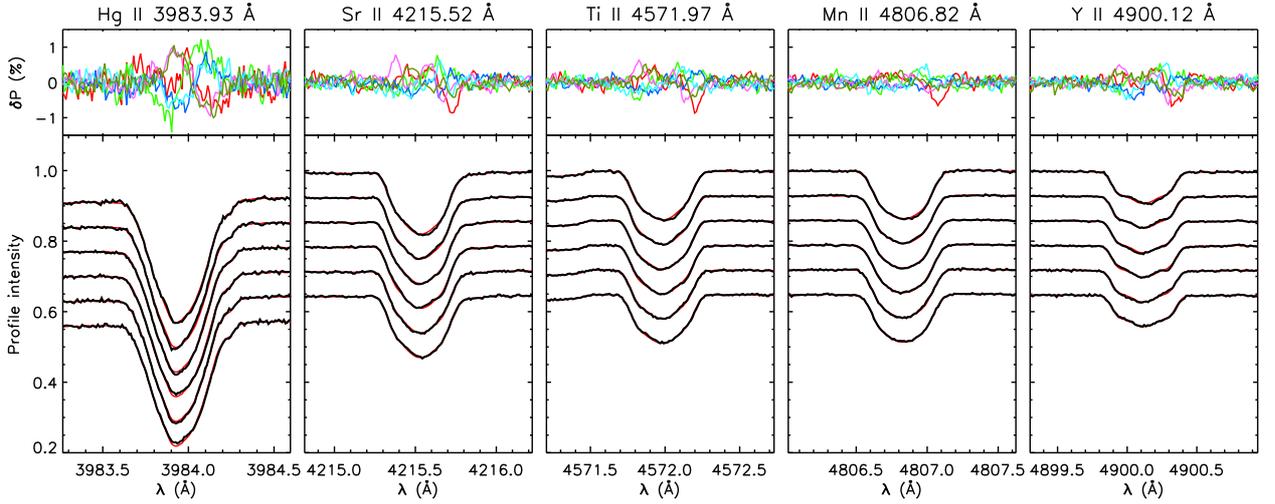}
\caption{Variability of individual \ion{Hg}{ii}, \ion{Sr}{ii}, \ion{Ti}{ii}, \ion{Mn}{ii}, and \ion{Y}{ii} spectral lines in \lep. The spectra corresponding to different observing nights are offset vertically. The mean profile (\textit{thin red line}) is plotted below the time-resolved spectra (\textit{thick black lines}). The upper part of each panel shows residual spectra.}
\label{fig:lpv}
\end{figure*}

\begin{figure*}[!t]
\centering
\fifps{16.5cm}{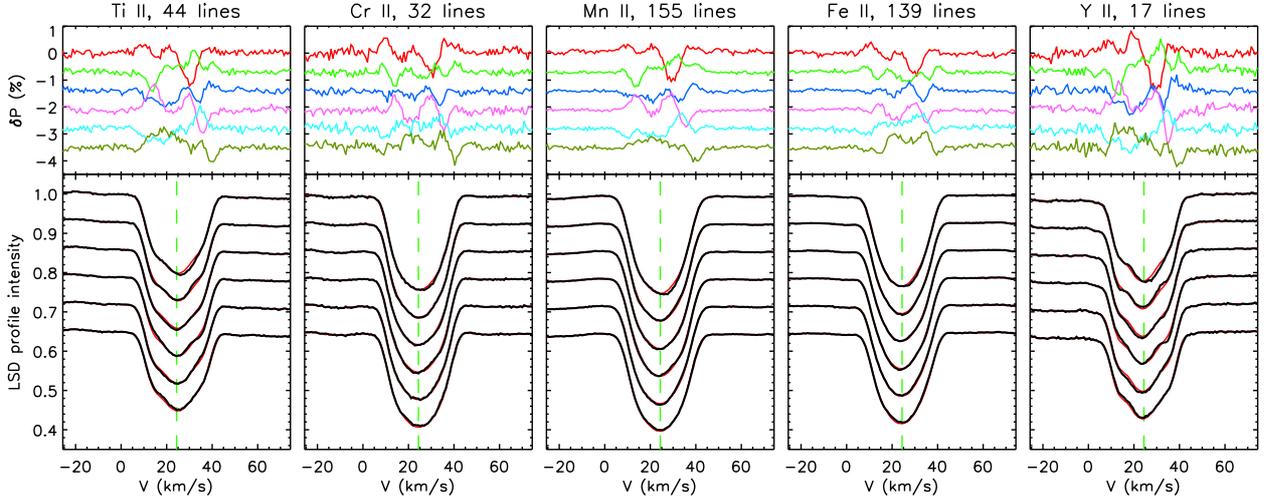}
\caption{Variability of the \ion{Ti}{ii}, \ion{Cr}{ii}, \ion{Mn}{ii}, \ion{Fe}{ii}, and \ion{Y}{ii} LSD profiles. The ion and the number of lines used for constructing mean profiles are indicated on top. The spectra corresponding to different observing nights are offset vertically. The upper panel in each column shows residual profile variation, while the bottom panel compares the time-averaged (\textit{thin red line}) and time-resolved (\textit{thick black lines}) LSD profiles. The vertical dashed line shows the mean stellar radial velocity.}
\label{fig:lpv_lsd}
\end{figure*}

We used the least-squares deconvolution (LSD) code and the methodology described by \citet{kochukhov:2010a} to perform a sensitive magnetic field search for \lep. This multiline technique assumes that each intensity or circular polarization line profile is a scaled copy of the common mean line shape and that overlapping lines add up linearly. Under these approximations, the high-quality average Stokes $I$ and $V$ profiles are derived from observations for a given set of spectral lines \citep{donati:1997}. 

The line mask necessary for applying LSD was extracted from the VALD database \citep{kupka:1999}, assuming stellar parameters $T_{\rm eff}$\,=\,12750~K and $\log g$\,=\,4.0 \citep[e.g.,][]{woolf:1999}, together with the abundances from \citet{adelman:2000b}. The final line mask, obtained after excluding the regions affected by the hydrogen lines and strong telluric features, included 526 spectral lines with a central intensity exceeding 10\% of the continuum. This list is dominated by \ion{Mn}{ii} and \ion{Fe}{ii} lines. The average wavelength and the mean effective Land\'e factor are 4769~\AA\ and 1.138, respectively. The application of LSD increases the S/N of the polarized profiles by a factor of $\approx$\,8 (see Table~\ref{tab1}).

The Stokes $I$ and $V$ LSD profiles for our four spectropolarimetric observations of \lep\ are presented in Fig.~\ref{fig:lsd}. These mean profiles were computed for the velocity range from $-70$ to +120~\kms, with a step of 0.8~\kms, which is close to the mean pixel spacing of HARPS spectra. 

There is no evidence of the magnetic signature in any of the LSD Stokes $V$ profiles. The false alarm probability assessment \citep[FAP,][]{donati:1992} indicates that none of the Stokes $V$ LSD spectra contains a statistically significant (FAP\,$<10^{-5}$) or even a marginal (FAP\,$<10^{-3}$) signal.

The mean longitudinal magnetic field, \bz, was estimated from the first moment of LSD Stokes $V$. The profiles were integrated within $\pm$\,22~\kms\ of the mean radial velocity of the star, $24.38\pm0.07$~\kms, estimated from LSD Stokes $I$. The \bz\ values range between $-2.5$ and 1.4~G with an error bar of 1.6--2.5~G. Individual measurements are reported in Table~\ref{tab1}. They are entirely consistent with the null hypothesis of no magnetic field. There is no detection of \bz\ in the null LSD profiles either, confirming the absence of spurious polarization. The weighted mean of all four \bz\ measurements is $0.39\pm0.93$~G, implying a 3$\sigma$ upper limit of 3~G for the longitudinal field.

\section{Line profile variability}
\label{lpv}

A low-amplitude variability is apparent in many individual metal lines in the spectrum of \lep. These changes do not exceed 1\% of the continuum intensity and require S/N\,$>$\,500 to be detected reliably. Figure~\ref{lpv} shows examples of the variability in individual lines of \ion{Hg}{ii}, \ion{Sr}{ii}, \ion{Ti}{ii}, \ion{Mn}{ii}, and \ion{Y}{ii}. The largest amplitude is seen for the \ion{Hg}{ii} 3984~\AA\ line. In addition, variability is evident in several lines of \ion{Sc}{ii}. Marginal changes are also present in numerous absorption features of \ion{Fe}{ii} and \ion{Cr}{ii}, as well as in a few strong lines of \ion{Si}{ii} and \ion{Mg}{ii}.

\begin{figure}[!th]
\centering
\fifps{8cm}{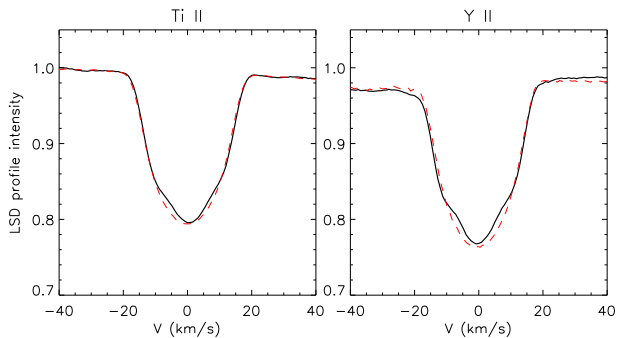}
\caption{Comparison of the time-averaged LSD profiles of \ion{Ti}{ii} and \ion{Y}{ii} derived from observations (\textit{solid black line}) and synthetic spectrum (\textit{dashed red line}).}
\label{fig:syn}
\end{figure}

Aiming to increase the precision of the line profile variability analysis, we applied the LSD technique to individual chemical elements following the multiprofile LSD approach introduced by \citet{kochukhov:2010a}. The resulting LSD profiles, derived from 17--155 lines, are shown in Fig.~\ref{fig:lpv_lsd}. The LSD analysis confirms variability in the \ion{Fe}{ii} and \ion{Cr}{ii} lines. 

The six HARPS spectra of \lep\ available to us are not sufficient for reliably determining the stellar rotational period. But the absence of a smooth progression of the residual profiles obtained for observations during consecutive nights in February 2011 suggests that rotational period does not exceed a few days. On the other hand, the stellar effective temperature $T_{\rm eff}$\,=\,12600$\pm200$~K and Hipparcos parallax $\pi$\,=\,$17.54\pm0.55$~mas \citep{van-leeuwen:2007} yield $R=3.39\pm0.16 R_{\sun}$. For this radius value and $v_{\rm e}\sin i$\,=\,$16\pm0.5$~\kms, the oblique rotator relation predicts a reasonable inclination angle of 30\degr--75\degr\ for the rotation period range of 5--10~d.

\section{Discussion}
\label{disc}

We have detected very low-amplitude line profile variations in the bright HgMn star \lep. In contrast to previous studies \citep{kochukhov:2005b,hubrig:2006,folsom:2010,makaganiuk:2011}, these changes were found not only for a few heavy elements but also for all Fe-peak ions represented in the stellar spectrum and, possibly, for the lines of light elements \ion{Si}{ii} and \ion{Mg}{ii}. 

The night-to-night changes of the residual profiles (Figs.~\ref{fig:lpv} and \ref{fig:lpv_lsd}) exhibit qualitatively similar, though not identical, line distortions for different elements. This indicates a similar inhomogeneous distribution of various elements. A more extended dataset is required to confirm this conclusion and rule out the possibility of a nonuniform temperature distribution.

The weak line profile variation in \lep\ suggests that low-contrast spots are ubiquitous in HgMn stars and that in many cases their signatures have been missed due to an insufficient precision of the spectroscopic data and lack of systematic time-resolved line profile studies.

Besides the low-amplitude variability discussed above, the time-averaged profiles of the \ion{Ti}{ii} and \ion{Y}{ii} lines exhibit anomalous triangular cores deviating significantly from the rotational Doppler line shape. This can be interpreted as a signature of a dominant axisymmetric component in the inhomogeneous abundance distribution with an enhancement of both elements at the rotational pole. The peculiar shapes of the average \ion{Ti}{ii} and \ion{Y}{ii} profiles is not an artifact of the LSD procedure since the LSD profiles derived from the synthetic spectrum using the same line mask do not show any anomaly (Fig.~\ref{fig:syn}).

Spectropolarimetric observations of \lep\ analyzed in our study yield the most precise magnetic field measurements for a spotted HgMn star. It is remarkable that even after reaching the precision of a few G we cannot detect longitudinal magnetic field. The lack of a statistically significant signal in the LSD Stokes $V$ profiles obtained during four different nights also rules out the presence of complex fields similar to those detected in cool active stars with the same instrument, data reduction, and analysis procedures as applied for \lep\ \citep{kochukhov:2011}.

The upper limit of 3~G for the longitudinal magnetic field obtained for \lep\ is much smaller than the equipartition field of $\approx$\,100~G expected in the line-forming region of the stellar atmosphere. Thus, it is unlikely that any field below a few G, if exists at all, can influence the accumulation of chemical elements by the atomic diffusion and govern the spot formation process. Time dependence of the radiative diffusion \citep{alecian:1998} or mixing induced by the hydrodynamical instabilities and circulation are more likely candidates for the mechanism behind the spot formation and evolution in HgMn stars. The similarity of the spot topologies found for different elements in 66~Eri \citep{makaganiuk:2011}, $\varphi$~Phe (Makaganiuk et al., submitted), and tentatively suggested here for \lep\ probably favors the latter hypothesis because diffusion instabilities must produce more diverse surface distributions due to their strong dependence on the radiative acceleration of individual chemical elements.

\begin{acknowledgements}
OK is a Royal Swedish Academy of Sciences Research Fellow, supported by grants from Knut and Alice Wallenberg Foundation and Swedish Research Council.
\end{acknowledgements}

\end{document}